\input harvmac.tex
%
\figno=0
\def\fig#1#2#3{
\par\begingroup\parindent=0pt\leftskip=1cm\rightskip=1cm\parindent=0pt
\baselineskip=11pt
\global\advance\figno by 1
\midinsert
\epsfxsize=#3
\centerline{\epsfbox{#2}}
\vskip 12pt
{\bf Fig. \the\figno:} #1\par
\endinsert\endgroup\par
}
\def\figlabel#1{\xdef#1{\the\figno}}
\def\encadremath#1{\vbox{\hrule\hbox{\vrule\kern8pt\vbox{\kern8pt
\hbox{$\displaystyle #1$}\kern8pt}
\kern8pt\vrule}\hrule}}

\overfullrule=0pt

%
\def\cqg#1#2#3{{\it Class. Quantum Grav.} {\bf #1} (#2) #3}
\def\np#1#2#3{{\it Nucl. Phys.} {\bf B#1} (#2) #3}

\def\prl#1#2#3{{\it Phys. Rev. Lett.}{\bf #1} (#2) #3}
\def\physrev#1#2#3{{\it Phys. Rev.} {\bf D#1} (#2) #3}

\def\cmp#1#2#3{{\it Comm. Math. Phys.} {\bf #1} (#2) #3}

\font\zfont = cmss10 

\def\bigone{\hbox{1\kern -.23em {\rm l}}}
\def\ZZ{\hbox{\zfont Z\kern-.4emZ}}

\def\a{\alpha}
\def\b{\beta}

\def\d{\delta}

\def\m{\mu}
\def\n{\nu}

\def\r{\rho}

\def\s{\sigma}

\def\ps{\psi}
\def\G{\Gamma}

\def\O{\Omega}

\Title{
{\vbox{
\rightline{\hbox{hepth/0011114}}
\rightline{\hbox{CALT-68-2305}}
}}}
{\vbox{
\hbox{\centerline{
A Note on Compactifications on }}
\hbox{\centerline{$Spin(7)$-Holonomy Manifolds}}
}}
\smallskip
\centerline{Katrin Becker\footnote{$^1$}
{\tt beckerk@theory.caltech.edu}} 
\smallskip
\centerline{\it California Institute of Technology 452-48, 
Pasadena, CA 91125}
\centerline{\it CIT-USC Center for Theoretical Physics}

\bigskip 
\bigskip

In this note we consider compactifications
of ${\cal M}$-theory on $Spin(7)$-holonomy manifolds to three-dimensional Minkowski 
space. In these compactifications a warp factor is included. 
The conditions for unbroken $N=1$ supersymmetry give rise to determining
equations for the 4-form field strength in terms of the warp factor
and the self-dual 4-form of the internal manifold.

\bigskip
\baselineskip 18pt
\bigskip
\noindent

\Date{November, 2000}

Warped compactifications of ${\cal M}$-theory and ${\cal F}$-theory have 
attracted recently much attention in connection to confining gauge
theories 
\ref\ps{J.~Polchinski and M.~J.~Strassler, ``The String Dual of a Confining
Four-Dimensional Gauge theory'', hep-th/0003136. }, 
\ref\kt{I.~R.~Klebanov and A.~A.~Tseytlin, ``Gravity Duals of Supersymmetric
$SU(N) \times SU(N+M)$ Gauge Theories'', \np {578} {2000} {123}, 
hep-th/0002159.},
\ref\ks{I.~R.~Klebanov and M.~J.~Strassler, ``Supergravity and a Confining
Gauge Theory: Duality Cascades and $\chi$SB-Resolution 
of Naked Singularities'',
JHEP {\bf 0008} (052) 2000, hep-th/0007191.},
\ref\gp{M.~Gra\~na and J.~Polchinski, ``Supersymmetric 3-Form Flux and
Perturbations of $AdS(5)$'', hep-th/0009211.}, 
\ref\gub{S.~S.~Gubser, ``Supersymmetry and F-Theory Realization 
of the Deformed Conifold with 3-Form Flux'', hep-th/0010010. } and the 
string theoretic realization of the Randall-Sundrum scenario 
\ref\rs{L.~Randall and R.~Sundrum, ``A Large Mass 
Hierarchy from a Small Extra
Dimension'', \prl {83} {1999} {3370}, hep-th/9905221.}, 
\ref\rs{L.~Randall and R.~Sundrum, ``An Alternative to 
Compactification'', 
\prl {83} {1999} {4690}, hep-th/9906064.} suggested in
\ref\cpv{C.~S.~Chan, P.~L.~Paul and H.~Verlinde,``A Note on 
Warped String Compactification'', \np {581} {2000}  {156}, hep-th/0003236.}.
In \ref\bb{K.~Becker and M.~Becker, ``${\cal M}$-Theory on Eight Manifolds'', 
\np {477} {1996} {155}, 
hep-th/9605053.} the conditions for unbroken supersymmetry for 
compactifications on Calabi-Yau 4-folds were found.
These are manifolds 
that admit two covariantly constant spinors.
It has recently been shown that the model considered by 
Klebanov and Strassler {\ks} can be obtained from the general type
of solutions presented in {\bb}.

From this perspective it is rather interesting to understand the 
physics of warped compactifications of ${\cal M}$-theory
\foot{Warped compactifications on 4-folds have been  
further considered in
\ref\gvw{S.~Gukov, C.~Vafa and E.~Witten, ``CFT'S From Calabi-Yau 
Four-Folds'', \np {584} {2000} {69}, hep-th/9906070.}, 
\ref\sethi{K. Dasgupta, G. Rajesh and S. Sethi, 
``M-Theory, Orientifolds and G-Flux'', JHEP {\bf 9908} (1999) 023, 
hep-th/9908088. }, 
\ref\gukov{S.~Gukov, ``Solitons, Superpotentials and Calibrations'',
\np {574} {2000} {169}, hep-th/9911011.} and  
\ref\ggw{S.~J.~Gates, S.~Gukov and E.~Witten, 
``Two Two-Dimensional Supergravity Theories from Calabi-Yau 
Four-Folds'', \np {584} {2000} {109}, hep-th/0005120.}.}.
In this note we would like to describe compactifications 
of ${\cal M}$-theory 
on $Spin(7)$-holonomy manifolds. 
These are eight-manifolds that admit only one covariantly 
constant Majorana-Weyl spinor which arises from the 
the decomposition $8_c \rightarrow 7 \oplus 1$. Therefore these 
compactifications will give rise to $N=1$ theories in three dimensions
while the models considered in {\bb} had an $N=2$ supersymmetry. 
These theories are interesting because they cannot be obtained 
from a compactification of any supersymmetric four-dimensional 
theory on an $S^1$. In fact, a theory with $N=1$ in $d=4$ would 
yield an $N=2$ theory in three-dimensions
\ref\vafa{C.~Vafa, ``Evidence for F-Theory'', \np {469} {1996} {403},
hep-th/9602022.}.

$Spin(7)$-holonomy manifolds can be treated in a similar way as the  
Calabi-Yau 4-fold case considered in {\bb} so we will be brief here and 
use the notations and conventions of {\bb}. 

To derive the conditions following from unbroken supersymmetry
we start with the supersymmetry transformation of the gravitino 
in eleven-dimensional supergravity
\eqn\aii{
\d \Psi_M  = \nabla_M\eta -{1 \over 288} 
({\G_M}^{PQRS}-8 \d_M^P \Gamma^{QRS} )F_{PQRS} \eta=0.
}
We make the following ansatz for the metric
\eqn\aiii{
g_{MN}(x,y)=
\Delta^{-1}(y) \pmatrix{
 g_{\m \n}(x) & 0 \cr
0 &  g_{mn}(y)\cr
}, 
}
where $x$ are the coordinates of the external space labeled by the indices
$\mu,\nu,\dots$ and $y$ are the coordinates 
of the internal manifold labeled by $m,n,\dots$. 
$\Delta=\Delta(y)$ is the warp factor.

The eleven-dimensional spinor $\eta$ is decomposed as
\eqn\aiv{
\eta=\epsilon \otimes \xi, 
}
where $\epsilon$ is a three-dimensional anticommuting spinor and 
$\xi$ is an eight-dimensional Majorana-Weyl spinor. Furthermore we
make the decomposition of the gamma matrices 
\eqn\dxi{
\eqalign{
\Gamma_{\mu}=\gamma_{\mu} \otimes \gamma_9,  \cr
\Gamma_m= 1 \otimes \gamma_m , \cr
}
}
where $\gamma_{\mu}$ and $\gamma_m$ are the gamma matrices of the 
external and internal space respectively. We choose the matrices 
$\gamma_m$ to be real and antisymmetric. $\gamma_9$ is the eight-dimensional 
chirality operator which anti-commutes with all the $\gamma_m$'s. 

In compactifications with maximally symmetric three-dimensional 
space-time the non-vanishing components of the 4-form field strength 
$F_4$ are
\eqn\av{
\eqalign{&  F_{mnpq}\quad {\rm arbitrary} \cr
& F_{\mu \nu \rho m} =\epsilon_{\m\n\r} f_m , \cr}
}
where $F_{mnpq}$ and  $f_m$  
will be determined from the conditions following from unbroken 
supersymmetry and $\epsilon_{\m\n\r}$ is the Levi-Civita 
tensor of the three-dimensional external space.
The external component of the gravitino supersymmetry transformation 
is given by the following expression
\eqn\avi{
\eqalign{ 
\delta \psi_{\mu} =\nabla_{\mu} \eta & -{1 \over 288} \Delta^{3/2} 
( \gamma_{\mu} \otimes  \gamma^{mnpq}) F_{mnpq} \eta \cr 
& + {1 \over 6} \Delta^{3/2} ( \gamma_\mu  \otimes \gamma^m ) 
f_m \eta\cr
 & +{1 \over 4} \partial_n ( \log \Delta) 
( \gamma_{\mu} \otimes \gamma^n ) \eta,  \cr} 
}
where we have used a positive chirality eigenstate 
$\gamma_9 \xi=\xi$. Considering negative chirality spinors
corresponds to an eight-manifold with a reversed orientation
\ref\ip{C.~J.~Isham and C.~N.~Pope, ``Nowhere vanishing Spinors 
and Topological Obstruction to the Equivalence of the 
NSR and GS Superstrings'',
\cqg {5} {1988} {257}, C.~J.~Isham, C.~N.~Pope and N.~P.~Warner, 
``Nowhere-vanishing Spinors and Triality Rotations in 8-Manifolds'',
\cqg {5} {1988} {1297}.}. 
Since we would like to compactify ${\cal M}$-theory to 
three-dimensional Minkowski space we impose the condition
\eqn\avii{
\nabla_{\mu} \epsilon =0.
}
The external component of supersymmetry is then reduced to 
the equation 
\eqn\aviii{
T \xi=0 \qquad {\rm with} \qquad 
T= F_{mnpq}\gamma^{mnpq}- 48( f_n-\partial_n \Delta^{-3/2} ) \gamma^n .
}
Taking into account that $\xi$ is Weyl we conclude
that the external component
of $F_4$ can be expressed in terms of the warp factor 
\eqn\ax{
F_{\m\n\r m} =\epsilon_{\m\n\r} \partial_m \Delta^{-3/2},  
}
while the internal component of $F_4$ (which we denote by $F$) is 
constrained to satisfy
\eqn\aax{
F_{mnpq}\gamma^{mnpq} \xi=0.
}

The analysis of the internal components of supersymmetry can be performed as 
in {\bb}. We find that in terms of the transformed quantities
\eqn\bxi{
\eqalign{
& {\tilde g}_{mn} =\Delta^{-3/2} g_{mn} ,\cr
& {\tilde \xi} =\Delta^{1/4} \xi , \cr
}
}
the internal component of the gravitino transformation 
is given by 
\eqn\bxii{
{\tilde \nabla}_m {\tilde \xi} + 
{1 \over 24} \Delta^{-3/4}  F_m {\tilde \xi} =0.
}
The metric ${\tilde g}_{mn}$ describes the $Spin(7)$-holonomy manifold. 
These manifolds are Ricci flat and they admit one covariantly 
constant spinor which satisfies 
\eqn\bxiii{
{\tilde \nabla}_m {\tilde \xi}=0. 
}
Therefore we see that $F$ has to satisfy 
\eqn\bxiv{
F_{mnpq} {\tilde \gamma}^{npq}{\tilde \xi}=0. 
}
Note that the condition {\bxiv} is actually stronger than {\aax}\foot{
This equation has been noticed before in 
\ref\hr{S.~W.~Hawking and M.~M.~Taylor-Robinson, ``Bulk Charges
in Eleven Dimensions'', \physrev {58} {1998} {025006}, hep-th/9711042.}.}.
However it can be shown
that if $F$ is self-dual then {\bxiv} is equivalent to {\aax}. 

The proof that $F$ is self-dual goes as
follows. 
From {\bxiv} we obtain the equation 
\eqn\bxv{
F_{mnpq}{\tilde \xi}^T \{ {\tilde \gamma}^{npq}, 
{\tilde \gamma}_{abc} \} {\tilde \xi} =0. 
}
To further evaluate {\bxv}  we note that 
we can construct covariantly constant $p$-forms 
in terms of the eight-dimensional spinor ${\tilde \xi}$ 
\eqn\aii{
\omega_{a_1 \dots a_p} ={\tilde \xi}^T 
{\tilde \gamma_{a_1\dots a_p}}{\tilde \xi}.
}
Since ${\tilde \xi}$ is Majorana-Weyl {\aii} is non-zero only for 
$p=0,4$ or 8 (see
\ref\gpp{
G.~W.~Gibbons, D.~N.~Page and C.~N.~Pope, 
``Einstein Metrics on $S^3$, $R^3$ and $R^4$ Bundles'', 
\cmp {127} {529} {1990}.}). 
The $Spin(7)$ calibration is then given by the closed 
self-dual 4-form
\eqn\aii{
{ \Phi}_{mnpq } ={\tilde \xi}^T {\tilde \gamma}_{mnpq } {\tilde \xi}.
}
If we would have considered negative chirality spinors this form 
would be anti-self-dual
\ref\hs{J.~A.~Harvey and A.~Strominger, 
``Octonionic Superstring Solitons'',
\prl {66} {1991} {549}.}.
Taking this definition of the calibration into account and
{\bxv} we obtain
\eqn\bxvi{
F_{mnpq} = { 3\over 2} F_{ab m [n } { \Phi_{pq]} }^{ab}.
}
By considering the quantity 
\eqn\fxii{
F_{mnpq}{\tilde \xi}^T \{ {\tilde \gamma}_{abcd}, {\tilde \gamma}^{mnpq} \} 
{\tilde \xi}=0, 
}
we can derive the condition
\eqn\fxiii{
\star F_{mnpq}+F_{mnpq}=3F_{ab[mn} {\Phi_{pq]}}^{ab}, 
}
where by $\star$ we mean the Hodge dual with respect to 
the metric of the eight-dimensional internal space. 
Antisymmetrizing the right hand side of {\bxvi} over 
the indices $(mnpq)$ and comparing with {\fxiii} 
we see that $F$ satisfies the self-duality condition 
\eqn\fxiv{
F=\star F. 
}
This self-duality condition can also be obtained 
from the equation of motion of $F$ by using the explicit form 
of the external component of $F_4$ {\hr}.

Taking this self-duality condition into account, we next would like
to show that the constraint ${\bxiv}$ coming from the internal component
of supersymmetry is equivalent to the condition ${\aax}$. 
For this purpose a useful identity to consider is 
\eqn\zi{
F_m  F^m =-{1\over 8} {\cal F}^2 -3F_{mnpq}
\left(F^{mnpq} -\star F^{mnpq} \right), 
}
where ${\cal F}=F_{mnpq}{\tilde \gamma}^{mnpq}$. 
Since $F$ is self-dual 
the second term on the right hand side {\zi} vanishes. 
Equation {\zi} then implies that {\bxiv} is satisfied 
if and only if {\aax} is fulfilled.

To see the conditions imposed by {\aax} on our flux we note that
Fierz rearrangements imply
\eqn\zii{
F_{mnpq}{\tilde \gamma}^{mnpq}{\tilde \xi}
=F_{mnpq}\Phi^{mnpq}{\tilde \xi}-{F_{mnpr}} {\Phi_s}^{mnp} 
{\tilde \gamma}^{rs} 
{\tilde \xi}.
}
The condition for unbroken supersymmetry states 
that the left hand side of {\zii} vanishes. After multiplying this 
equation by ${\tilde \xi}^T$ we get  
\eqn\ziv{
F\wedge \Phi=0. 
}
Vanishing of the second term on the right hand side of {\zii} then implies 
\eqn\zzv{
\omega_{rs}{\tilde \gamma}^{rs}{\tilde \xi}=0, }
where we have defined a 2-form $\omega$ as
\eqn\ziii{
\omega=
{1\over 2} F_{mnpr} {\Phi_s}^{mnp} dx^r \wedge dx^s.
}
The spinors ${\tilde \gamma}^{rs}{\tilde \xi}$ are not independent
\ref\moore{M. Marino, R. Minasian, G. Moore and 
A. Strominger, ``Nonlinear Instantons from Supersymmetric P-Branes'',  
JHEP {\bf 0001} (005) 2000, hep-th/9911206.}. 
To satisfy {\zzv} $\omega$ has to obey the self-duality constraint
\eqn\zvi{
{1\over 2} \Phi^{rspq} \omega_{pq}=\lambda \omega^{rs}
}
with $\lambda=1$. But by taking the relation 
\eqn\zvii{
\Phi^{mnpt} \Phi_{qrst}=6\delta_{qrs}^{mnp}-9 \delta_{[q}^{[m}
{\Phi^{np]}}_{rs]}
}
and the definition of $\omega$ into account it is easy to see that 
$\omega$ satisfies {\zvi} with $\lambda=-3$. Therefore we conclude
\eqn\zv{
\omega=0.}
Equations {\fxiv}, {\ziv} and {\zv} are the 
determining equations for $F$. They are the necessary and 
sufficient conditions for {\bxiv} to be satisfied.

After imposing the self-duality condition {\fxiii} takes the form
\eqn\fxv{
F_{mnpq}={3\over 2} F_{ab[mn} {\Phi_{pq]}}^{ab}. 
}
By contracting this equation with $\Phi^{npqr}$ it is possible to 
show that {\fxv} is equivalent to {\ziv} and {\zv}. Therefore this is 
another way of expressing the condition for unbroken 
supersymmetry. In this form the condition for unbroken supersymmetry is 
similar to the one satisfied by Yang-Mills fields 
${\cal F}_{mn}$ for which the following 
equivalence relation holds \ref\cdf{
E. Corrigan, C. Devchand and D. B. Fairlie, ``First-Order Equations 
for Gauge Fields in Spaces of Dimension Greater than Four'', 
\np {214} {1983} {452}. }
\eqn\fxvi{
{\cal F}_{mn}={1\over 2} \Phi_{mnpq} {\cal F}^{pq}
\qquad \Longleftrightarrow 
\qquad {\cal F}_{mn}{\tilde \gamma}^{mn }{\tilde \xi}=0.
}

The determining equation for the warp factor is the 
fivebrane Bianchi identity which after using our solution
for $F_4$ takes the same form as in {\bb}
\eqn\axi{
d \star d \log \Delta = { 1\over 3} F \wedge F -{ 2 \over 3} 
( 2 \pi)^4 X_8.  
} 

For compactifications of ${\cal M}$-theory on $Spin(7)$-holonomy manifolds  
we can expect to find non-vanishing expectation values for $F_4$ 
independently of the fact that the manifold is compact or not. 
This is different than the situation considered in 
\ref\bbfour{K. Becker and M. Becker, 
``Compactifying ${\cal M}$-Theory to Four Dimensions'', 
JHEP {\bf 0011} (2000) 029, hep-th/0010282.} 
in which compactifications of ${\cal M}$-theory on seven-dimensional 
manifolds were considered and which only had non-vanishing expectation 
values for $F$ in the case that the internal manifold was non-compact. 

For a compact $Spin(7)$-holonomy manifold $K^8$ we can integrate {\axi} 
and obtain the relation 
\eqn\axii{
\int_{K^8} F \wedge F + { 1\over 12} \chi=0, 
}
where $\chi$ is the Euler number of the eight-manifold {\bb}, 
\ref\svw{S.~Sethi, C.~Vafa and E.~Witten, ``Constraints on Low Dimensional
String Compactification'', \np {480} {1996} {213}, hep-th/9606122.}. 

To summarize we have found the following conditions for unbroken 
supersymmetry for compactifications of ${\cal M}$-theory on manifolds with 
$Spin(7)$ holonomy 
to three-dimensional Minkowski space: the internal components of $F_4$ 
obey the constraints {\fxiv}, {\ziv} and {\zv}, 
the external components of $F_4$ are 
determined in terms of the warp factor by {\ax} and the warp factor can 
then be obtained from equation {\axi}.

Concrete examples of compact $Spin(7)$-holonomy manifolds 
have been constructed in 
\ref\joyce{D. Joyce, ``Compact 8-Manifolds with Holonomy $Spin(7)$'', 
{\it Invent. Math.} {\bf 123} (1996) 507.} (see also
\ref\acha{B. S. Acharya, ``On Mirror Symmetry for Manifolds 
of Exceptional Holonomy'', \np {524} {1998} {269}, hep-th/9707186; 
B. S. Acharya, J. M. Figueroa-O'Farril, C. M. Hull and 
B. Spence, ``Branes at Conical Singularities and Holography'', 
{\it Adv. Theor. Math. Phys.}, {\bf 2} (1999) 1249.} for further discussion). 
A non-compact example was discussed recently in \ref\clp{
M. Cvetic, H. Lu and C. N. Pope, ``Brane Resolution Through 
Transgression'', hep-th/0011023.}\foot{See also {\hr}.} and was 
constructed in {\gpp}. 
In this case the explicit form of the metric is known 
and it takes the form of a quaternionic line bundle over a 4-sphere:
\eqn\cxi{
ds_8^2=\a(r)^2dr^2+\b(r)^2(\s^i-A^i)^2+\gamma(r)^2 d\O_4^2, 
}
where $\s_i$ are left-invariant 1-forms of $SU(2)$, $A^i$
are $SU(2)$ Yang-Mills potentials on the unit 4-sphere whose 
metric is $d\O_4^2$. We will be following the notation of 
{\clp} and refer the reader to this work for further details. 

In {\clp} an explicit computation of $F$ was done. In the following 
we would like to show that this solution satisfies our equations. 
This solution is anti-self-dual. 
This would correspond to choosing spinors 
with negative chirality (i.e. which 
satisfy $\gamma_9 {\tilde \xi}=-{\tilde \xi}$)
instead of the positive chirality spinors that we have used here. 
To show that equation {\ziv} is satisfied we need 
the explicit forms of $\Phi$ and $F$. They are given by
\eqn\cxii{
\Phi=f_1 'dr \wedge \epsilon_{(3)}-(f_1+g_1) Y_{(4)} +g_1' dr \wedge X_3 
-6 g_1 \O_{(4)}, 
}
with 
\eqn\cxiii{
f_1={ 1\over 5} c_1 (1-6z) z^{-6/5} 
\qquad {\rm and }\qquad g_1=c_1 z^{-6/5} , 
}
while $F$ is given by
\eqn\cxiv{
F=f_2 'dr \wedge \epsilon_{(3)}-(f_2+g_2) 
Y_{(4)} +g_2' dr \wedge X_3 
-6 g_2 \O_{(4)},
}
with 
\eqn\cxv{
f_2={ 1\over 5} (z-6) z^{1/5} \qquad {\rm and} \qquad g_2=z^{1/5} . 
}
Using this explicit forms for $F$ and $\Phi$ it is easy to see that 
$F \wedge \Phi$ is proportional to the quantity
\eqn\cxvi{
g_1 f'_2+g_2 f'_1+g_1'(f_2+g_2)+g_2' (f_1+g_1),
}
which vanishes after using {\cxiii} and {\cxv}. 
In the same way it is possible to show that the 2-form 
$\omega$ vanishes. This is easily seen in the orthonormal basis 
${\hat e}^I$ introduced in {\gpp}.

A superpotential in terms of the calibration describing
compactifications on $Spin(7)$-holonomy manifolds has been conjectured in
{\gukov}. It would be interesting to see if the conditions for unbroken 
supersymmetry obtained in 
this paper can actually be derived from the superpotential presented in
{\gukov}.

\vskip 1cm

\noindent {\bf Acknowledgment}

\noindent 

I thank M. Becker, S. Gukov, J. Polchinski and E. Witten for 
useful discussions and correspondence. 
This work was supported by the U.S. Department of Energy 
under grant DE-FG03-92-ER40701.

\listrefs

\end